\documentclass[%
 reprint,
superscriptaddress,
 amsmath,amssymb,
 aps,
]{revtex4-2}

\usepackage{setspace}
\usepackage{placeins} 
\usepackage{graphicx}
\usepackage{dcolumn}
\usepackage{bm}
\usepackage{hyperref}
\hypersetup{colorlinks=true, pdfstartview=FitV, linkcolor=blue, citecolor=blue, plainpages=false, pdfpagelabels=true, urlcolor=blue}

\begin{document}

\preprint{APS/123-QED}

\title{Determination of the Spacing Between Hydrogen-Intercalated Quasi-Free-Standing Monolayer Graphene and 6H-SiC(0001) Using Total-Reflection High-Energy Positron Diffraction}

\author{Matthias Dodenh\"oft}
\affiliation{Forschungs-Neutronenquelle Heinz Maier-Leibnitz FRM II, Technische Universit\"at M\"unchen, Lichtenbergstraße 1, 85748 Garching, Germany}

\author{Izumi Mochizuki}
\author{Ken Wada}
\author{Toshio Hyodo}
\affiliation{Institute of Materials Structure Science, High Energy Accelerator Research Organization (KEK), Ibaraki 305-0801, Japan}

\author{Peter Richter}
\author{Philip Sch\"adlich}
\author{Thomas Seyller}
\affiliation{Institute for Physics, Technische Universit\"at Chemnitz, Reichenhainer Straße 70, 09126 Chemnitz, Germany}
\affiliation{Center for Materials, Architectures and Integration of Nanomembranes (MAIN), Rosenbergstraße 6, 09126 Chemnitz, Germany}

\author{Christoph Hugenschmidt}
 \email{Christoph.Hugenschmidt@frm2.tum.de}
\affiliation{Forschungs-Neutronenquelle Heinz Maier-Leibnitz FRM II, Technische Universit\"at M\"unchen, Lichtenbergstraße 1, 85748 Garching, Germany}

\date{\today}


\begin{abstract}

We have investigated the structure of hydrogen-intercalated quasi-free-standing monolayer graphene (QFMLG) grown on 6H-SiC(0001) by employing total-reflection high-energy positron diffraction (TRHEPD). At least nine diffraction spots of the zeroth order Laue zone were resolved along \mbox{$<$11$\bar{2}$0$>$} and three along \mbox{$<$1$\bar{1}$00$>$}, which are assigned to graphene, SiC and higher order spots from multiple diffraction on both lattices. We further performed rocking curve analysis based on the full dynamical diffraction theory to precisely determine the spacing between QFMLG and the SiC substrate. Our study yields a spacing of \mbox{$d_{\mathrm{QFMLG}} = 4.18(6)\thinspace$\r{A}} that is in excellent agreement with the results from density-functional theory (DFT) calculations published previously.

\end{abstract}


\maketitle

\section{\label{sec:Intro}Introduction}

Graphene has been extensively studied due to its exceptional properties, such as extremely high thermal conductivity \cite{Bal08} and mechanical strength \cite{Lee08, Cao20}, as well as massless charge carriers with unconventional behavior in tunneling, confinement or magnetotransport \mbox{\cite{Nov05, Kat06, Cas09}}. Among different approaches to produce large-area graphene on an industrial scale, its synthesis on the surface of the wide band gap semiconductor SiC \cite{Hee07, Emt09, Kru16} is particularly appealing for high-power or high-frequency electronics. This is based on the fact that epitaxial graphene can be directly grown on SiC(0001) (without transfer), which is an intrinsic technological advantage.

Upon thermal decomposition, SiC(0001) exhibits a characteristic \mbox{($6 \sqrt{3} \times 6 \sqrt{3}$)$R 30^{\circ}$} surface reconstruction, which can be regarded as a precursor to the growth of graphene layers \cite{Bom75, For98, Sta09}. The associated buffer layer consists of carbon atoms and is covalently bound to the SiC substrate. Therefore, it has a corrugated structure \cite{Var08, Gol13} and is electrically inactive \cite{Rie09}. On the other hand, this interface also deteriorates the electronic properties of the adjacent graphene layer, leading to an intrinsic n-doping, a long-range corrugation in the density of states and a reduced, temperature-dependent charge carrier mobility \cite{Oht07, Rie10b, Job10}.

Effective decoupling from the SiC substrate can be achieved by hydrogen intercalation, as first demonstrated by \mbox{Riedl \emph{et al.} \cite{Rie09}}. At temperatures of \mbox{$\sim \! 500^{\circ}$C}, hydrogen can migrate below the buffer layer and break the covalent bonds with the substrate. The buffer layer is converted into quasi-free-standing monolayer graphene (QFMLG), which resides on the \mbox{H-terminated} SiC(0001) surface \cite{Spe11, Spe17}. Although there are other possibilities of interface manipulation, such as rapid cooling \cite{Bao16} or the intercalation of various other elements \mbox{\cite{Gie10, Wal11, For16, Ben21, Ryb21}}, hydrogen intercalation is considered to be the best approach for obtaining high-quality graphene \cite{Sak22}.

In this study, we investigated the surface structure of hydrogen-intercalated QFMLG on 6H-SiC(0001) using total-reflection high-energy positron diffraction (TRHEPD) \cite{Ich92, Kaw98, Fuk19, Hug16}, i.e. positron diffraction in grazing incidence. In contrast to electrons, positrons experience a repulsive crystal potential that leads to an outstanding surface sensitivity. As demonstrated by Fukaya \mbox{\emph{et al.}}, TRHEPD is particularly well suited to analyze the potential buckling of 2D materials \cite{Fuk13b, Fuk16b} and precisely determine the spacing between graphene and different substrates \cite{Fuk16a}. In an early study, \mbox{Kawasuso \emph{et al.}} already investigated the surface graphitization and structure of few-layer graphene on \mbox{6H-SiC(0001)} \cite{Kaw05}. However, since then the theoretical understanding of the material system and the sample preparation has improved significantly, as well as the positron beam quality available for TRHEPD. Recently, \mbox{Endo \emph{et al.}} investigated pristine and Ca-intercalated bilayer graphene grown on \mbox{6H-SiC(0001) \cite{End20}}. To the best of our knowledge, the present study is hence the first analysis of hydrogen-intercalated QFMLG using TRHEPD.

\section{\label{sec:experiment}Experimental procedure}

As a prerequisite, TRHEPD requires a bright and coherent positron beam of adequate intensity. The measurements were therefore carried out at the Slow Positron Facility (SPF) of the Institute of Materials Structure Science, KEK in Japan \cite{Wad12, Mae14}. Currently, this is the only operational TRHEPD setup in the world, although a second setup at the NEutron induced POsitron source \mbox{MUniCh} (\mbox{NEPOMUC}) \cite{Hug12, Hug14} has been developed recently \cite{Dod21}. A comprehensive overview of the measurement technique TRHEPD can be found elsewhere \cite{Fuk19, Hug16}.

The QFMLG sample was prepared by polymer-assisted sublimation growth (PASG) of a buffer layer on 6H-SiC(0001) and subsequent hydrogen intercalation \cite{Kru16}. Prior to TRHEPD, the sample was comprehensively pre-characterized using X-ray photoelectron spectroscopy (XPS), atomic force microscopy (AFM) and low-energy electron diffraction (LEED). This confirmed the presence of QFMLG and allows to exclude contaminations. Further details on the preparation parameters and the pre-characterization can be found in the supplementary notes \ref{sec:sample_preparation} and \ref{sec:pre-characterization}, respectively.

After transfer to the TRHEPD ultra-high vacuum (UHV) chamber at the SPF, the QFMLG sample was annealed in-situ at $500^{\circ}$C to remove surface adsorbates. To ensure a clean surface, we explicitly compared the effect of different annealing temperatures and durations on the basis of the obtained TRHEPD rocking curves (see supplementary note \ref{sec:heat_treatment}). All TRHEPD measurements were conducted at room temperature and we employed reflection high-energy electron diffraction (RHEED) for qualitative comparison. The incident, brightness-enhanced positron beam was set to an energy of 10$\thinspace$keV. Diffraction patterns were recorded along the azimuthal directions \mbox{$<$11$\bar{2}$0$>$} and  \mbox{$<$1$\bar{1}$00$>$} (many-beam condition), as well as \mbox{$7.5^{\circ}$} off the high-symmetry directions (one-beam condition). In order to obtain the experimental rocking curves, i.e. the intensity of the specular spot as a function of the glancing angle $\theta$, the sample was tilted in steps of 0.1$^{\circ}$ with respect to the incident positron beam. Following a standard routine, the intensity of the specular spot was extracted from the TRHEPD patterns as outlined in supplementary note \ref{sec:data_treatment}.

\begin{figure} [!thb]
\centering
\includegraphics[trim = 0mm 0mm 0mm 0mm, clip, width=8.7cm]{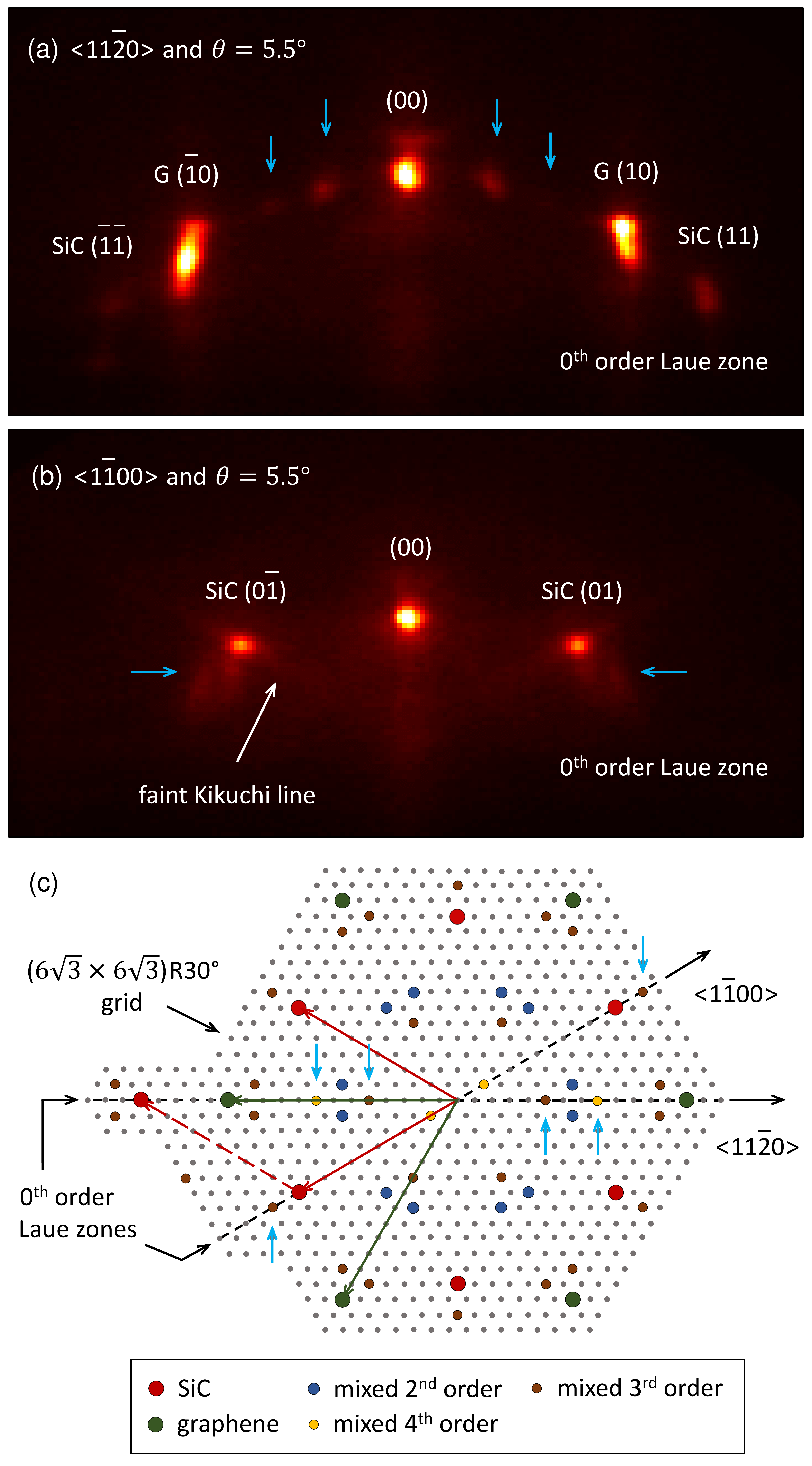}
\setlength\belowcaptionskip{0pt}
\setlength{\abovecaptionskip}{0pt plus 2pt minus 2pt}
\caption{\label{fig:TRHEPD_QFMLG_diffraction_pattern}TRHEPD patterns of QFMLG along high-symmetry directions (integration time 40$\thinspace$min, linear color scale). (a) The bright diffraction spots along \mbox{$<$11$\bar{2}$0$>$} are assigned to the graphene lattice. Only the zeroth order Laue zone is resolved. (b) The two pronounced spots along \mbox{$<$1$\bar{1}$00$>$} are assigned to the SiC lattice. Faint Kikuchi lines can be observed as well. (c) Reciprocal space: \mbox{($6 \sqrt{3} \times 6 \sqrt{3}$)$\thinspace R 30^{\circ}$} grid with SiC (red) and graphene (green) lattice vectors and points. Higher order spots that stem from multiple diffraction on both lattices are marked as well (not all fourth order spots are shown). The blue arrows indicate the mixed higher order spots that are assigned in (a) and (b), respectively.}
\end{figure}

\section{\label{sec:results}Results and discussion}

First, the positron diffraction patterns are discussed qualitatively by assignment of the observed diffraction spots. Subsequently, we employ TRHEPD rocking curve analysis to determine the interlayer spacing between QFMLG and the SiC substrate.

\subsection{\label{subsec:TRHEPD_pattern}Positron diffraction patterns}

Two TRHEPD patterns for $\theta=5.5^{\circ}$ are shown in \mbox{figure \ref{fig:TRHEPD_QFMLG_diffraction_pattern}(a)} \mbox{and (b)}. We observe at least nine diffraction spots along \mbox{$<$11$\bar{2}$0$>$} and three along \mbox{$<$1$\bar{1}$00$>$}. Additionally, we can identify very faint Kikuchi lines, which are much less pronounced than in RHEED. This is related to the shallower probing depth of TRHEPD due to the repulsive crystal potential for \mbox{positrons \cite{Fuk14}}.

Following the LEED analysis, we expect the TRHEPD patterns to be composed of diffraction spots from graphene and SiC, which are superimposed by further spots that stem from multiple diffraction on both lattices. Since the lattices of graphene and SiC are rotated by 30$^{\circ}$ with respect to each other, the main diffraction spots are observed along different crystallographic directions. Due to the different lattice constants, the spacing of the graphene spots on the Laue semicircle must be greater than for SiC \cite{Bos16}. Therefore, we can assign the two bright spots in \mbox{figure \ref{fig:TRHEPD_QFMLG_diffraction_pattern}(a)} to graphene and those in \mbox{figure \ref{fig:TRHEPD_QFMLG_diffraction_pattern}(b)} to SiC. The ratio of the respective horizontal spacings is approximately 1.29, which agrees reasonably well with the calculated inverse ratio of the lattice constants of $\sim \! 1.25$. Interestingly, we find that the intensity of the graphene spots is much larger than those of SiC, although only the topmost surface layer is composed of graphene. This is explained by the exceptional surface sensitivity of TRHEPD, even at such large glancing angle. For comparison, we refer to the LEED pattern shown in figure \mbox{S3(a)} of the supplementary material, where the SiC spots are more pronounced indicating a deeper mean probing depth.

For the assignment of the remaining diffraction spots, we refer to \mbox{figure \ref{fig:TRHEPD_QFMLG_diffraction_pattern}(c)}, which is a map of reciprocal space including (mixed) higher order diffraction spots. The blue arrows indicate the relevant mixed order spots that coincide with the zeroth order Laue zones. For the measurement along \mbox{$<$11$\bar{2}$0$>$}, all four spots are clearly resolved and the relative intensities of third and fourth order support this assignment. With respect to the SiC lattice, these spots are located at \mbox{$\pm$(5/18, 5/18)} and \mbox{$\pm$(8/18, 8/18)}, respectively. The two minor diffraction spots further outside on the Laue semicircle in \mbox{figure \ref{fig:TRHEPD_QFMLG_diffraction_pattern}(a)} can be assigned to the (11) and ($\bar{1} \bar{1}$) reflections of SiC. For the measurement along \mbox{$<$1$\bar{1}$00$>$}, we expect to see the two mixed third order spots close to the SiC (01) and (0$\bar{1}$) reflections. However, these spots are kind of smeared out in the diffraction pattern and it is challenging to determine their exact location.

\begin{figure} [!htb]
\centering
\includegraphics[trim = 0mm 0mm 0mm 0mm, clip, width=8.0cm]{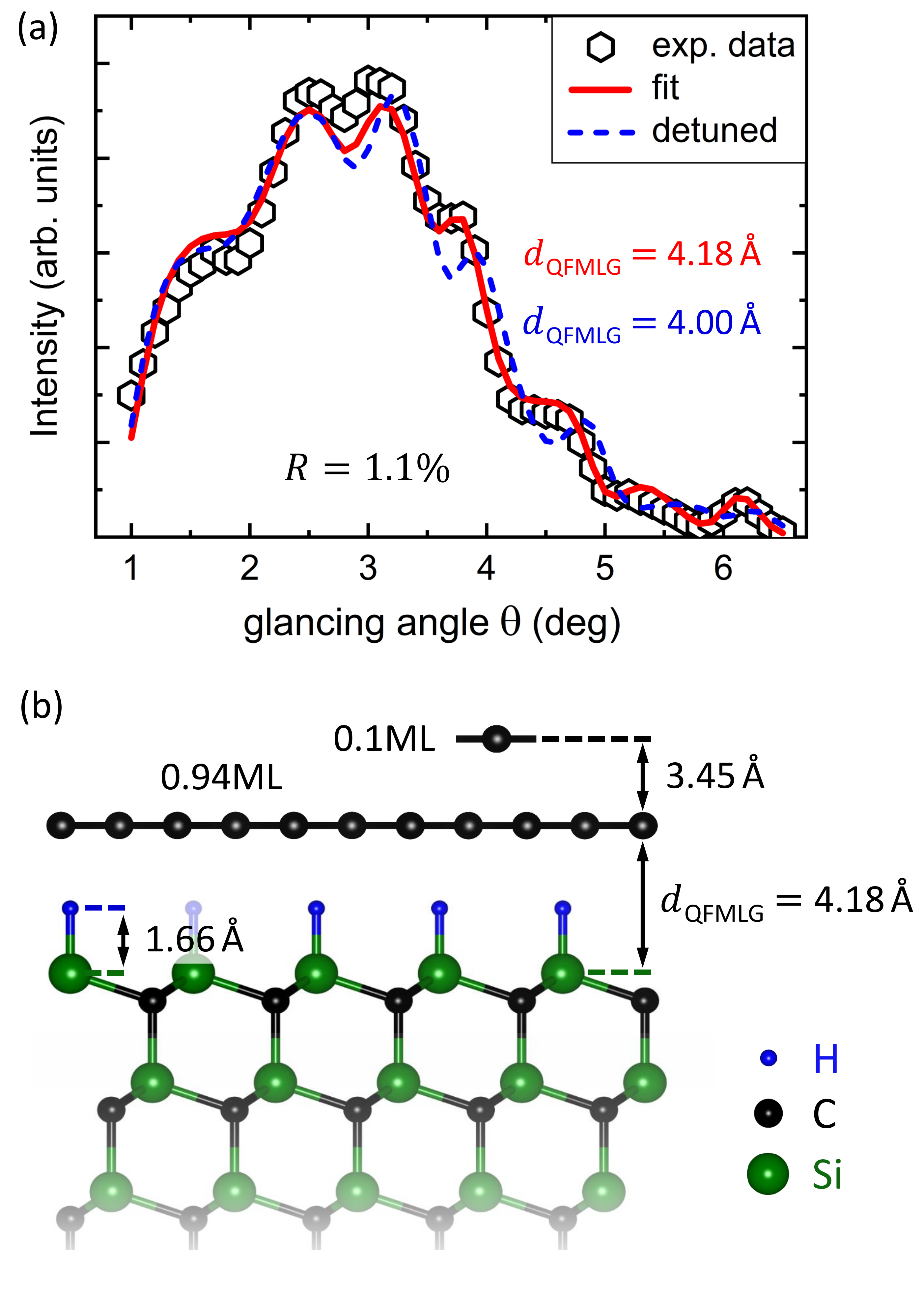}
\setlength\belowcaptionskip{0pt}
\setlength{\abovecaptionskip}{0pt plus 2pt minus 2pt}
\caption{\label{fig:Final_Fit_QFMLG_OB} TRHEPD one-beam rocking curve analysis of QFMLG. The calculation is based on the structure depicted in (b), including six fit parameters, i.e. occupations and spacings of QFMLG and graphene bilayer, the spacing of the hydrogen atoms and the lattice relaxation of the top Si layer. The structural parameters were extracted from the fit in (a) (red line). The rocking curve is particularly sensitive to the spacing $d_{\mathrm{QFMLG}}$: when detuned by 0.18$\thinspace$\r{A} (dashed blue line), the $R$-factor increases from 1.1$\%$ to 1.7$\%$ and some features change completely.}
\end{figure}

\subsection{\label{subsec:rocking_curve_analysis_one-beam}Rocking curve analysis}

The intensities of the TRHEPD rocking curves were calculated numerically based on the full dynamical
diffraction theory \cite{Ich83, Han95, Tan20}. To account for the finite size of the sample, the calculated intensities were scaled by the geometrical factor \mbox{$\sin{\theta}$}. The average crystal potential was set to 11.5$\thinspace$eV for graphene \cite{Fuk16a}, whereas the value for SiC was fine-tuned to 17.1$\thinspace$eV to match the experimentally observed Bragg peaks \cite{Kaw03}. We employed the Nelder-Mead algorithm to minimize the difference between the calculated and experimental rocking curves by adjusting the parameters of the established structure model \cite{Sfo15}. The agreement of the fit was quantified using the reliability factor \cite{Fuk19}
\begin{equation}
R =  \sqrt{ \thinspace \sum_{i} \left[ I_{\thinspace \mathrm{exp}}(\theta_{i}) - I_{\thinspace \mathrm{cal}}( \theta_{i}) \right]^2},
\label{eq:reliability_factor}
\end{equation}
where $I_{\thinspace \mathrm{exp}}(\theta_{i})$ and $I_{\thinspace \mathrm{cal}}(\theta_{i})$ are the normalized experimental and calculated intensities of the specular spot at the glancing angles $\theta_{i}$.

The open symbols in figure \mbox{\ref{fig:Final_Fit_QFMLG_OB}(a)} represent the experimental rocking curve obtained under the one-beam condition where the contribution of in-plane diffraction is strongly suppressed \cite{Ich87}. The one-beam analysis thus allows the determination of the out-of-plane atomic coordinates and yields the separation of atomic layers when including their mean composition and occupation \cite{Ich04}. The calculations were first carried out for a simple model assuming a fully occupied graphene layer without surface roughness or bilayer fraction. However, this model failed to reproduce the shoulder observed in the region of total reflection at $1.6^{\circ}$. At the same position, \mbox{Kawasuso \emph{et al.}} observed a pronounced dip structure, which was essentially attributed to atomic-scale surface roughness \mbox{\cite{Ich92, Kaw00c, Kaw03}}. We found that the shoulder can be well reproduced by including a small fraction of bilayer graphene. The best fit is the solid red line shown in figure \mbox{\ref{fig:Final_Fit_QFMLG_OB}(a)}, which yields a reliability factor of 1.1$\%$. The precisely known bulk lattice parameters of SiC were fixed, but we considered the possible relaxation of the uppermost Si layer. Since the flat structure of QFMLG has already been confirmed by other studies \cite{Sfo15}, we neglected buckling in the final fit to prevent overfitting. The results from the fit are listed in \mbox{table \ref{tab:Results_structure_QFMLG}} and the associated structure is schematically depicted in figure \mbox{\ref{fig:Final_Fit_QFMLG_OB}(b)}.

\begin{table*}
\caption{\label{tab:table3}Structural parameters of QFMLG extracted from the fit shown in \mbox{figure \ref{fig:Final_Fit_QFMLG_OB}(a)}. The spacings $d_{\mathrm{BL}}$, $d_{\mathrm{QFMLG}}$ and $d_{\mathrm{H}}$ are defined with respect to the position of the top Si layer (relaxed \mbox{by $\Delta \thinspace z_{\mathrm{Si}}$}).}
\label{tab:Results_structure_QFMLG}
\begin{ruledtabular}
\begin{tabular}{cccccc}
\multicolumn{2}{c}{bilayer fraction} & \multicolumn{2}{c}{QFMLG} & hydrogen & top Si layer\\
\hline
occupation & $d_{\mathrm{BL}}$ (\r{A}) & occupation & $d_{\mathrm{QFMLG}}$ (\r{A}) & $d_{\mathrm{H}}$ (\r{A}) & $\Delta \thinspace z_{\mathrm{Si}}$ (\r{A})\\
\hline
9.8$\%$ & 7.63$\thinspace \pm \thinspace$0.20 & 93.9$\%$ & 4.18$\thinspace \pm \thinspace$0.06 & 1.66$\thinspace \pm \thinspace$0.25 & $-$0.01$\thinspace \pm \thinspace$0.05\\
\end{tabular}
\end{ruledtabular}
\end{table*}

We emphasize that the rocking curve is particularly sensitive to the parameter $d_{\mathrm{QFMLG}}$, i.e. the spacing between QFMLG and the SiC substrate. This can be illustrated by deliberately detuning $d_{\mathrm{QFMLG}}$, as shown by the dashed blue line in figure \mbox{\ref{fig:Final_Fit_QFMLG_OB}(a)}. When the optimum value is reduced by only \mbox{0.18$\thinspace$\r{A}}, several features change completely, e.g. the shoulder at $\sim \! 4.6^{\circ}$ becomes a minimum and the reliability factor increases to 1.7$\%$ (or at best to 1.5$\%$ by re-adjusting all other parameters). Conversely, the strong influence on the rocking curve allows us to determine $d_{\mathrm{QFMLG}}$ with highest precision. In fact, already the first structure model (without bilayer occupation) yields very consistent results. In contrast, the uncertainties of the other parameters are significantly higher. The TRHEPD signal generally tends to be more sensitive to heavier atoms due to their increased atomic scattering factors \cite{Hos21}. In agreement with this, we observe that the variation of the spacing of the hydrogen layer has very little effect on the rocking curve. Consequently, the uncertainty is relatively high and other techniques, such as infrared spectroscopy, are better suited to precisely determine the Si-H bond length. Furthermore, we point out that the calculation did not explicitly consider bilayer domains, but individual C atoms that are distributed on the surface. In combination with the reduced occupation of the QFMLG layer, this can be interpreted as surface roughness that might originate from the sample preparation or intrinsic defects. The best fit yields a bilayer spacing of 3.45$\thinspace$\r{A} with respect to the QFMLG layer, which is slightly larger than the interlayer spacing in graphite \cite{Bas55}.

\begin{figure} [!thb]
\centering
\includegraphics[trim = 0mm 0mm 0mm 0mm, clip, width=7.4cm]{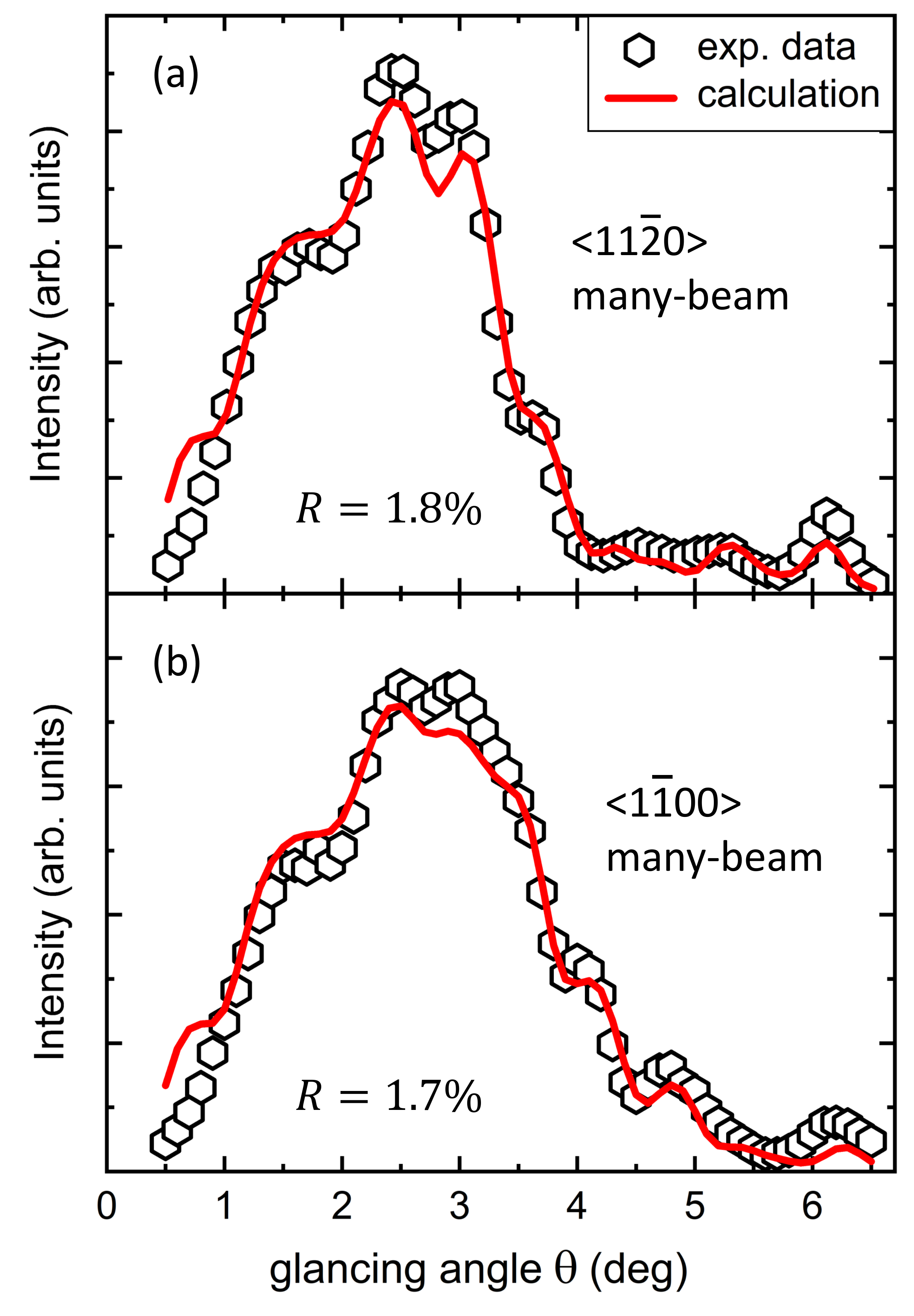}
\setlength\belowcaptionskip{0pt}
\setlength{\abovecaptionskip}{0pt plus 2pt minus 2pt}
\caption{\label{fig:Fit_many_beam_QFMLG} TRHEPD many-beam rocking curve analysis of QFMLG along (a) \mbox{$<$11$\bar{2}$0$>$} and (b) \mbox{$<$1$\bar{1}$00$>$}. The calculations were done for the structural parameters obtained from the one-beam analysis (see table \ref{tab:Results_structure_QFMLG}) and the literature values for the in-plane coordinates of graphene and SiC.}
\end{figure}

The rocking curves obtained under the many-beam condition along \mbox{$<$11$\bar{2}$0$>$} and \mbox{$<$1$\bar{1}$00$>$} are shown in figure \mbox{\ref{fig:Fit_many_beam_QFMLG}(a)} and (b), respectively. Since the exact in-plane coordinates of pristine graphene are known and the interaction with the SiC substrate is relatively weak, the many-beam data was mainly used for a sanity check of the obtained results. The calculations were thus performed for fixed parameters, i.e. without employing the Nelder-Mead algorithm. For both directions we obtain good agreement with reliability factors \mbox{of 1.8$\%$} \mbox{and 1.7$\%$}, respectively. We observe that the calculated curves deviate from the experimental data for very small glancing angles below 1$^{\circ}$ due to the significant decrease in intensity for geometrical reasons. We find that the shoulder at $\sim \! 1.6^{\circ}$ is well reproduced by the bilayer occupation determined before. Interestingly, the agreement in the angular range above $4^{\circ}$ is noticeably worse than for the one-beam data. This is most likely related to the presence of different stacking terminations of 6H-SiC(0001), as associated with terrace steps of different heights \cite{Sta04, Pak20}. Since step bunching is not dominant for PASG, the signal from different stacking terminations with distinct in-plane structures is superimposed. For the one-beam analysis, this has essentially no effect because the vertical spacings are the same for all stacking orders. However, it affects the many-beam rocking curve, particularly with increasing probing depth, i.e. for large glancing angles. In the calculations, on the other hand, we only considered one specific stacking termination due to computational constraints. Regardless of this, altogether the many-beam data supports the findings from the one-beam analysis, e.g. the presence of a small fraction of bilayer graphene associated with surface roughness and the found interlayer spacing $d_{\mathrm{QFMLG}}$.

\mbox{Sforzini \emph{et al.}} performed DFT calculations of the full supercell of QFMLG on \mbox{6H-SiC(0001)} within the generalized gradient approximation (GGA) and by employing the Perdew-Burke-Ernzerhof (PBE) functional with a correction for van der Waals effects \cite{Sfo15}. Our experimental result of $d_{\mathrm{QFMLG, \thinspace TRHEPD}} = (4.18 \pm 0.06$)$\thinspace$\r{A} is in excellent agreement with the calculated value of $d_{\mathrm{QFMLG, \thinspace DFT}} = 4.16 \thinspace$\r{A}, highlighting the strength of TRHEPD rocking curve analysis. In contrast, other studies based on normal incidence X-ray standing wave (NIXSW) \cite{Sfo15} or high-resolution X-ray reflectivity (XRR) measurements \cite{Eme14} yield a marginally larger spacing of $d_{\mathrm{QFMLG, \thinspace NIXSW}} = (4.22 \pm 0.06) \thinspace$\r{A} which coincides with theory just within the uncertainty range. Moreover, we note that the obtained surface relaxations of the substrate layers are particularly large in the NIXSW study, which is inconsistent.

\section{\label{sec:conclusion}Conclusion}

We performed TRHEPD measurements to investigate the surface structure of hydrogen-intercalated QFMLG epitaxially grown on 6H-SiC(0001). The observed diffraction spots were assigned to graphene, SiC and higher order spots that stem from multiple diffraction on both lattices. In contrast to LEED and RHEED, the graphene spots in the TRHEPD patterns were found to be much brighter than those of SiC due to the outstanding surface sensitivity of positron diffraction. For the quantitative analysis, we compared the experimental rocking curves with those calculated from the established structure model. Under the one-beam condition, the rocking curve of QFMLG was found to be particularly sensitive to the spacing between graphene layer and SiC substrate. Compared to previous experimental studies based on NIXSW or XRR, we find a slightly smaller spacing of \mbox{$d_{\mathrm{QFMLG}} = (4.18 \pm 0.06$)$\thinspace$\r{A}}, which is in excellent agreement with the value predicted by DFT calculations.

\begin{acknowledgments}
We thank Florian Speck for his support and discussions. Financial support by the German federal ministry of education and research (BMBF) within the project no. 05K16WO7 is gratefully acknowledged.
\end{acknowledgments}


\FloatBarrier

\bibliographystyle{apsrev4-2}
\bibliography{bibliography_TRHEPD_QFMLG_paper}

\pagebreak
\widetext
\begin{center}
\textbf{\large -- Supplemental Materials --\\Determination of the Spacing Between Hydrogen-Intercalated Quasi-Free-Standing Monolayer Graphene and 6H-SiC(0001) Using Total-Reflection High-Energy Positron Diffraction}
\end{center}
\setcounter{equation}{0}
\setcounter{figure}{0}
\setcounter{table}{0}
\setcounter{section}{0}
\setcounter{page}{1}
\makeatletter
\renewcommand{\theequation}{S\arabic{equation}}
\renewcommand{\thefigure}{S\arabic{figure}}
\renewcommand{\thetable}{S\arabic{table}}
\renewcommand{\thepage}{S\arabic{page}} 
\renewcommand{\thesection}{S\arabic{section}}  


In these supplementary notes, we provide information on the sample preparation, pre-characterization and data processing. Sample preparation parameters are presented in supplementary note \ref{sec:sample_preparation}. The results of the characterization prior to total-reflection high-energy positron diffraction (TRHEPD) are briefly discussed in supplementary note \ref{sec:pre-characterization}. After transfer to the TRHEPD ultra-high vacuum (UHV) chamber, surface adsorbates have been removed by in-situ heat treatment, which is evaluated in supplementary note \ref{sec:heat_treatment}. The procedure to extract experimental rocking curves from a set of individual positron diffraction patterns is explained in supplementary note \ref{sec:data_treatment}.

\section{\label{sec:sample_preparation} Sample details and preparation parameters}

A nitrogen \mbox{n-doped} \mbox{6H-SiC(0001)} wafer from the company PAM-Xiamen was cut along the \mbox{$<$1$\bar{1}$00$>$} and \mbox{$<$11$\bar{2}$0$>$} directions to obtain the $4.8 \times 10 \times 0.33 \thinspace$mm$^{3}$ sized substrate. The Si-terminated surface has been finished by an industrial-grade chemical-mechanical polishing (CMP) procedure, leading to a marginal surface roughness of less than 0.5$\thinspace$nm. Additional hydrogen etching, as reported in other studies \mbox{(e.g. \citep{Spe17})}, was therefore not necessary. 

The buffer layer was synthesized using the polymer-assisted sublimation growth (PASG) technique \cite{Kru16}, i.e. the sample was coated with polymer adsorbates, which serve as an initial source of carbon. For this, the photoresist \emph{AZ5214E} was used. After rinsing in isopropyl alcohol, the excess polymer was removed with a spin coater (100$\thinspace$rps). Subsequently, the sample was successively annealed at different temperatures and gas environments to induce polymer decomposition, buffer layer growth and hydrogen intercalation in order to obtain quasi-free-standing monolayer graphene (QFMLG). The respective process steps are summarized in table \ref{annealing_SiC_wafer}.

\vspace{1mm}

\begin{table}[h]
\centering
\begin{tabular}{|c|c|c|c|c|}
\hline
step & atmosphere & $T$ ($^{\circ}$C) & $t$ (min) & purpose \\
\hline
1 & UHV & 900 & $\thinspace$ 30 $\thinspace$ & $\thinspace$ polymer decomposition $\thinspace$ \\
\hline
2 & $\thinspace$ Ar (1$\thinspace$bar) $\thinspace$ & 1200 & 10 & annealing \\
\hline
3 & Ar (1$\thinspace$bar) & 1400 & 5 & growth of buffer layer \\
\hline
4 & $\thinspace$ $\mathrm{H_{2}}$ (880$\thinspace$mbar) $\thinspace$ & $\thinspace$ 550 $\thinspace$ & 90 & hydrogen intercalation  \\
\hline
\end{tabular}
\setlength\belowcaptionskip{0pt}
\setlength{\abovecaptionskip}{8pt plus 2pt minus 2pt}
\caption{Process parameters for PASG and hydrogen intercalation. The temperature ramp was set to 3$\thinspace$K/s and the sample cooled down naturally after step 1 and 3, respectively. During the hydrogen intercalation in step 4, the $\mathrm{H_{2}}$ flow rate was set to \mbox{15.2$\thinspace$mbar$\thinspace \cdot \thinspace$l/s.}}
\label{annealing_SiC_wafer}
\end{table}

\vspace{-3mm}

\section{\label{sec:pre-characterization}Sample pre-characterization}

We employed X-ray photoelectron spectroscopy (XPS), atomic force microscopy (AFM) and low-energy electron diffraction (LEED) to evaluate the sample quality, topography and surface structure, respectively.

\subsection*{\label{sec:XPS_Graphene}X-ray photoelectron spectroscopy}

\begin{figure} [!htb]
\centering
\includegraphics[trim = 0mm 0mm 0mm 0mm, clip, width=15.0cm]{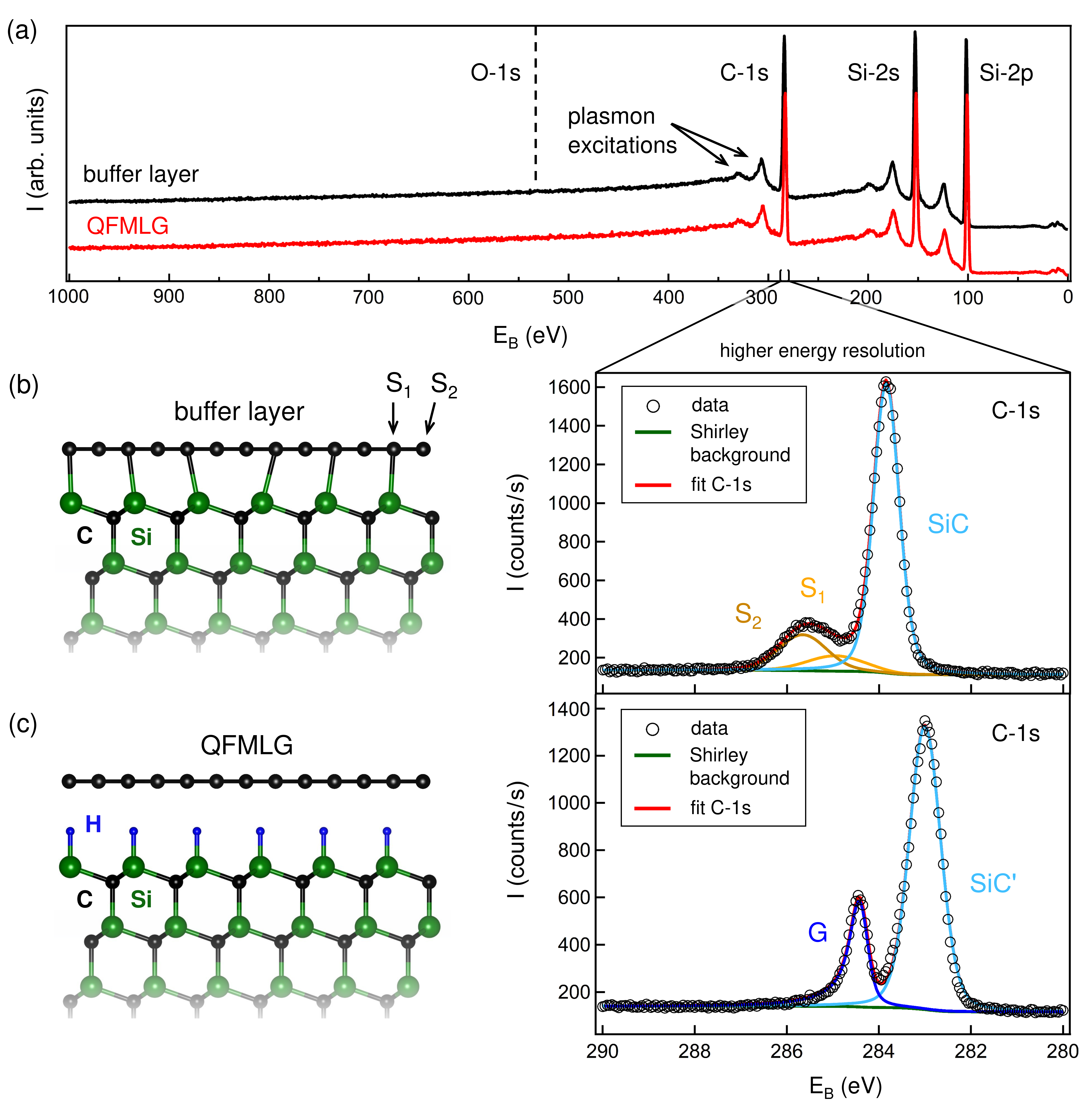}
\setlength\belowcaptionskip{0pt}
\setlength{\abovecaptionskip}{1pt plus 2pt minus 2pt}
\caption{\label{fig:XPS_data}XPS characterization before and after hydrogen intercalation. (a) The two spectra are shifted vertically for clarity. The three distinct peaks are related to C or Si core levels and there are no indications of contaminations. (b), (c) Left: illustration of the respective structure (schematic). Right: related high-resolution XPS spectra of the \mbox{C-1s} peak. The indicated components were fitted by a standard routine considering a Voigt line shape and a Shirley background. (b) The components $\mathrm{S_{1}}$ and $\mathrm{S_{2}}$ can be assigned to the buffer layer, while the large peak corresponds to the C atoms of the substrate \cite{Emt08}. (c) For QFMLG, the buffer layer components $\mathrm{S_{1}}$ and $\mathrm{S_{2}}$ are absent.}
\end{figure}

X-ray photoelectron spectroscopy (XPS) was applied to exclude possible contaminations. \mbox{Figure \ref{fig:XPS_data}(a)} shows the XPS spectra of the buffer layer and the QFMLG sample, i.e. before and after hydrogen intercalation. The main peaks of both spectra are assigned to the \mbox{carbon 1s} and the \mbox{silicon 2s} \mbox{and 2p} core levels. Apart from the adjacent minor peaks that stem from plasmon excitations, we do not observe additional peaks. Consequently, there are no significant contaminations and we can particularly exclude the partial intercalation of oxygen \cite{XPS_21}.

Further information can be obtained from the chemical shifts \mbox{of $E_{\mathrm{B}}$} within the \mbox{C-1s} peak. The spectrum of the buffer layer in \mbox{figure \ref{fig:XPS_data}}(b) exhibits a large peak with a side lobe at higher binding energy and can be fitted with three components. Most of the signal stems from the C atoms in the substrate while the side lobe is composed of the components $\mathrm{S_{1}}$ and $\mathrm{S_{2}}$ that are assigned to the buffer \mbox{layer \cite{Emt08}}. As shown in \mbox{figure \ref{fig:XPS_data}(c)}, the XPS spectrum of QFMLG can be well fitted by two components without contribution from $\mathrm{S_{1}}$ and $\mathrm{S_{2}}$. We thus conclude that the hydrogen intercalation successfully converted the buffer layer into QFMLG.

\subsection*{\label{sec:AFM_results}Atomic force microscopy}

\begin{figure} [!htb]
\centering
\includegraphics[trim = 0mm 0mm 0mm 0mm, clip, width=17.0cm]{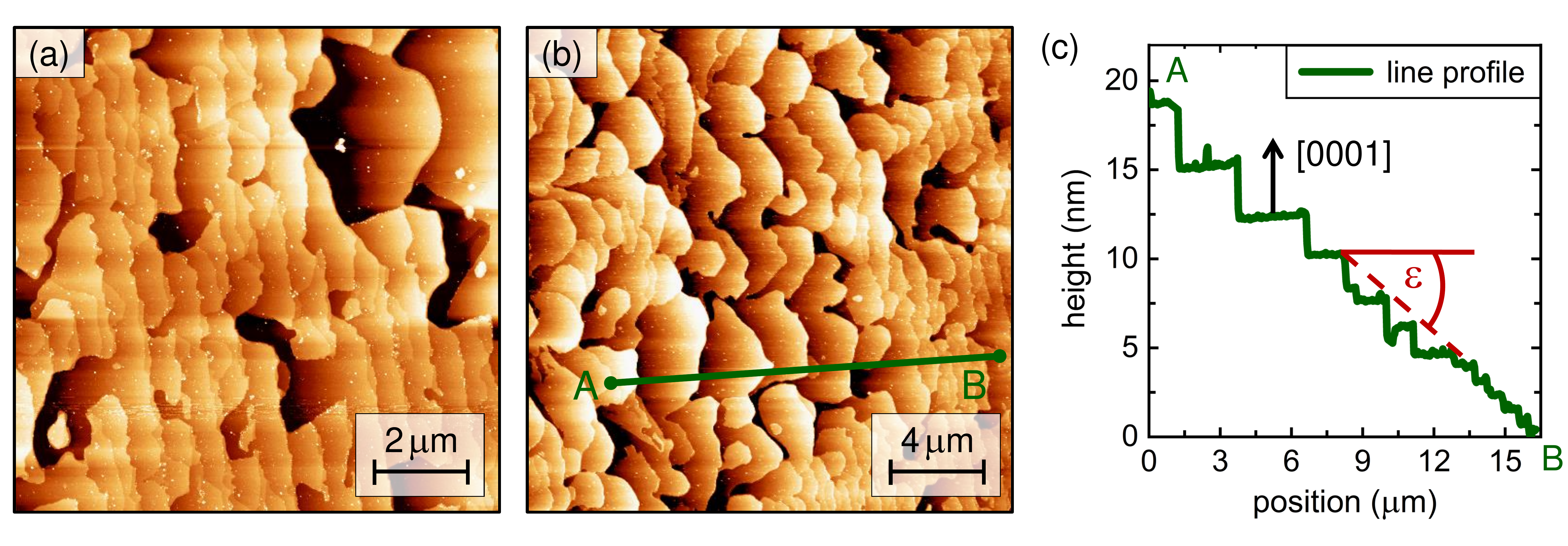}
\setlength\belowcaptionskip{0pt}
\setlength{\abovecaptionskip}{2pt plus 2pt minus 2pt}
\caption{\label{fig:AFM_characterization}The AFM images of the QFMLG sample in (a) and (b) indicate a terrace structure with various step heights. (c) This is confirmed by the line profile along the green path, which reveals larger steps when approaching point A. The wafer miscut angle is calculated to $\varepsilon \approx 0.07^{\circ}$.}
\end{figure}

Two AFM images of the QFMLG sample are shown in \mbox{figure \ref{fig:AFM_characterization}(a)} and (b). We observe minor defects and irregular regions on a scale of $\mu$m, which might be caused by a not perfectly uniform polymer adsorption. Due to the finite manufacturing accuracy, on-axis aligned SiC wafers have a small miscut, which leads to a stepped terrace structure on the surface \cite{Sta04}. The line profile perpendicular to the terrace steps along the green path in \mbox{figure \ref{fig:AFM_characterization}(b)} is plotted \mbox{in (c)}. The step heights vary from relatively large steps of more than 2 SiC unit cells close to the point A, to smaller steps in region B that correspond to 2-3 SiC bilayers. The coalescence of several small steps into a single step of greater height is known as step bunching and is usually less pronounced for \mbox{PASG \cite{Kru16, Pak20}}. However, with regard to TRHEPD measurements, these larger steps are not considered problematic. The miscut angle of the SiC(0001) substrate was calculated to $\varepsilon \approx 0.07^{\circ}$ by evaluating four different line profiles.

\subsection*{\label{sec:LEED_results}Low-energy electron diffraction}

\begin{figure} [!htb]
\centering
\includegraphics[trim = 0mm 0mm 0mm 0mm, clip, width=16cm]{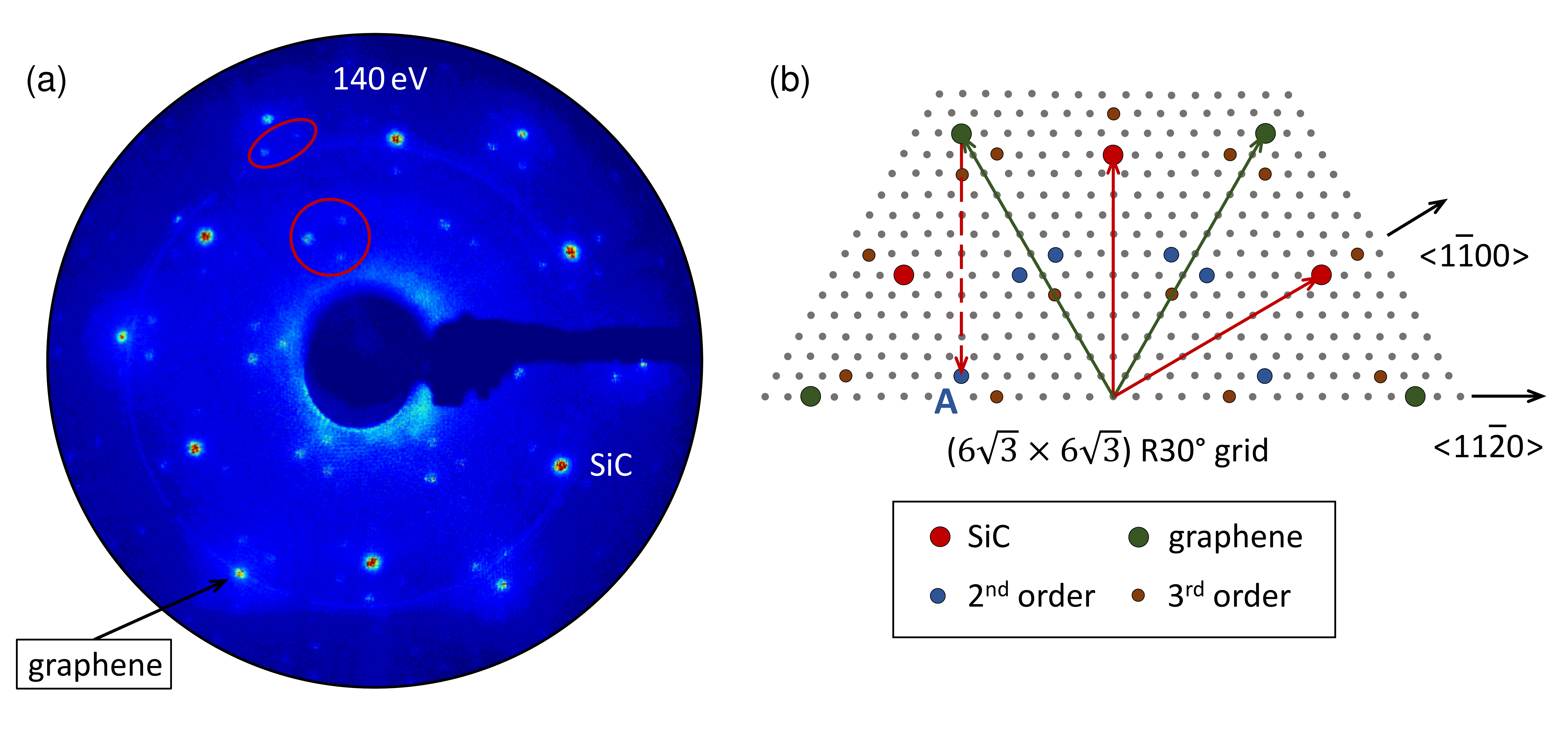}
\setlength\belowcaptionskip{0pt}
\setlength{\abovecaptionskip}{2pt plus 2pt minus 2pt}
\caption{\label{fig:LEED_characterization}LEED pattern of QFMLG. (a) We observe pronounced diffraction spots that are assigned to the SiC substrate and the graphene layer, respectively. Further spots of lower intensity are also present, e.g. the ones encircled in red. \mbox{(b) Reciprocal} lattice vectors of graphene (green) and SiC (red) on a \mbox{($6 \sqrt{3} \times 6 \sqrt{3}$)$\thinspace R 30^{\circ}$} grid. The locations of (mixed) higher order diffraction spots are obtained by linear combination of both lattice vectors, as illustrated for the second order \mbox{spot A}. When multiple diffraction on both lattices is considered, the additional spots observed in (a) can be well reproduced.}
\end{figure}

Qualitative LEED measurements were performed to analyze the surface structure. As shown in \mbox{figure \ref{fig:LEED_characterization}(a)}, we obtain a clear diffraction pattern and thus conclude that the sample has a long-range periodic order, which is a prerequisite for TRHEPD measurements. The individual diffraction spots are assigned to SiC and graphene, superimposed by additional spots that originate from multiple diffraction on both 2D lattices \cite{For98}. This is illustrated in \mbox{figure \ref{fig:LEED_characterization}(b)}. The first order diffraction spots are represented by green and red dots, respectively. By the linear combination of a graphene and a SiC reciprocal lattice vector (green and dashed red line) we reach the point A, which is thus a (mixed) second order diffraction spot. The faint diffraction spots encircled in red in \mbox{figure \ref{fig:LEED_characterization}(a)} are thus identified as second and third order spots. Since both pairs of lattice vectors match the \mbox{($6 \sqrt{3} \times 6 \sqrt{3}$)$\thinspace R 30^{\circ}$} grid, these higher order diffraction spots coincide with the location of fractional order spots as observed for the superstructure of a buffer layer.

\section{\label{sec:heat_treatment}Thermal stability of QFMLG}

Prior to quantitative TRHEPD measurements, the QFMLG sample has been annealed in-situ to remove adsorbates, which was necessary due to the exposure to air during transport. Since the hydrogen intercalation is known to be reversible, the annealing temperature must be sufficiently low \cite{Rie10b, Vir10b}. \mbox{Riedl \emph{et al.}} reported that QFMLG is stable up \mbox{to $700^{\circ}$C \cite{Rie09}}, which should however be regarded as an upper limit because it marks the onset of the Si-H bond \mbox{breaking \cite{Sie03}}. To exclude hydrogen desorption from the interface, we thus started with a rather low annealing temperature of $300^{\circ}$C and successively increased the temperature and duration. The cool-down time after each step was at least 30$\thinspace$min and all TRHEPD patterns were recorded at room temperature.

\begin{figure} [!htb]
\centering
\includegraphics[trim = 0mm 0mm 0mm 0mm, clip, width=14.9cm]{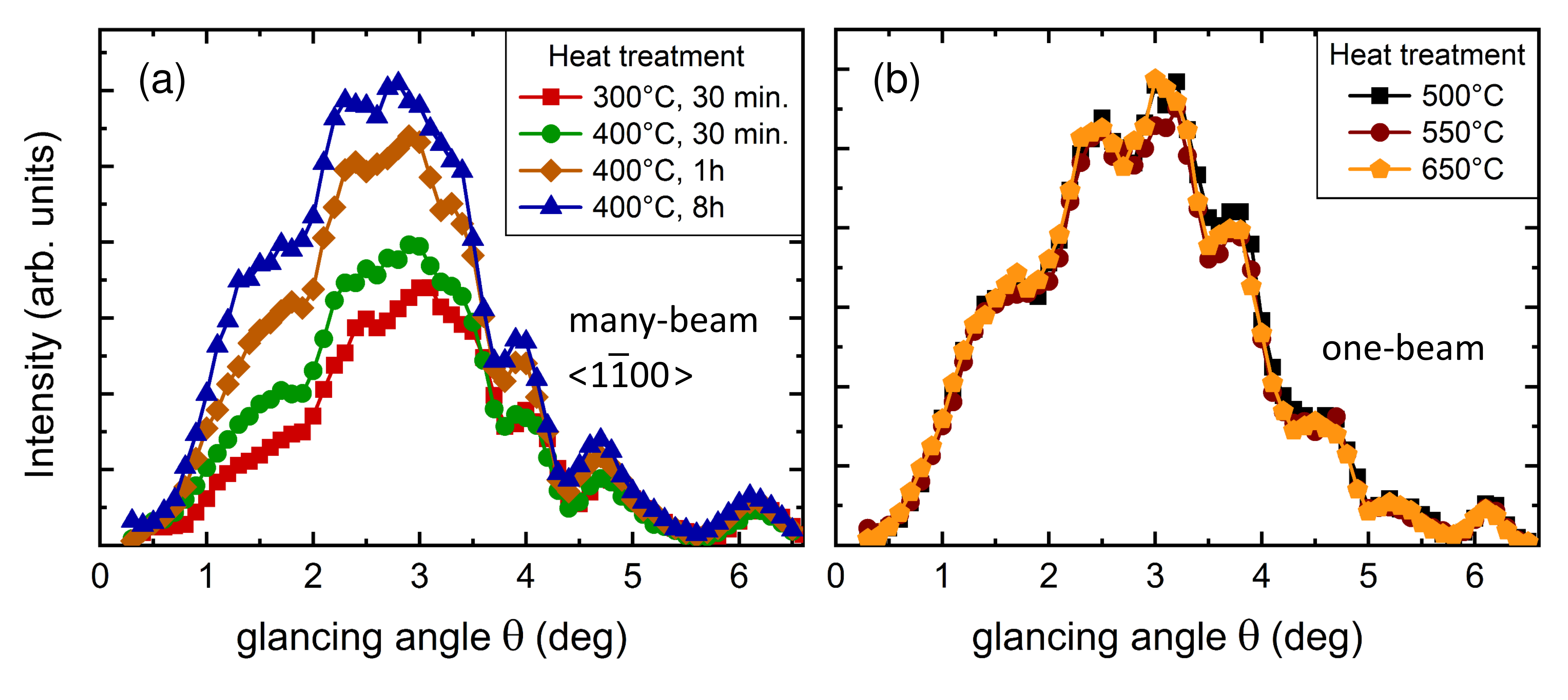}
\setlength\belowcaptionskip{0pt}
\setlength{\abovecaptionskip}{0pt plus 2pt minus 2pt}
\caption{\label{fig:Evaluation_heat_treament}Heat treatment of QFMLG. Intensities of the (00) spot were extracted from $7 \times 7$ pixel arrays (see \ref{sec:data_treatment}) and the rocking curves are not normalized. (a) The increase of intensity for $T_{\text{anneal}} \thinspace \! \leq \thinspace \! 400^{\circ}$C reveals that the surface is not completely adsorbate-free yet. (b) One-beam rocking curves after additional, step-wise annealing at higher temperatures (30$\thinspace$min each). No further increase in intensity is observed, i.e. we obtained a clean surface.}
\end{figure}

\mbox{Figure \ref{fig:Evaluation_heat_treament}(a)} shows the obtained rocking curves along \mbox{$<$1$\bar{1}$00$>$} (many-beam condition) and for $T_{\text{anneal}} \thinspace \! \leq \thinspace \! 400^{\circ}$C. The intensity of the specular spot increases steadily when extending the heat treatment, which holds true for the entire angular range. This is a clear indication that the surface is not fully conditioned yet. A clean surface can be obtained when annealing at $\sim 500^{\circ}$C, which is in agreement with the degassing procedure reported by \mbox{Emery \emph{et al.} \cite{Eme14}}. Upon further annealing up to temperatures of $650^{\circ}$C we observe no changes and the rocking curve can be well reproduced, as shown in \mbox{figure \ref{fig:Evaluation_heat_treament}(b)} for the one-beam condition. On the other hand, this also confirms that there is no hydrogen desorption, which would be accompanied by a structural change evident in the rocking curve. This is consistent with the results from \mbox{Riedl \emph{et al.}} \cite{Rie09}. For the quantitative analysis, we used the data recorded after annealing at $500^{\circ}$C, averaged over three separate rocking scans.

\section{\label{sec:data_treatment}TRHEPD data processing}

\begin{figure} [!htb]
\centering
\includegraphics[trim = 0mm 0mm 0mm 0mm, clip, width=18.0cm]{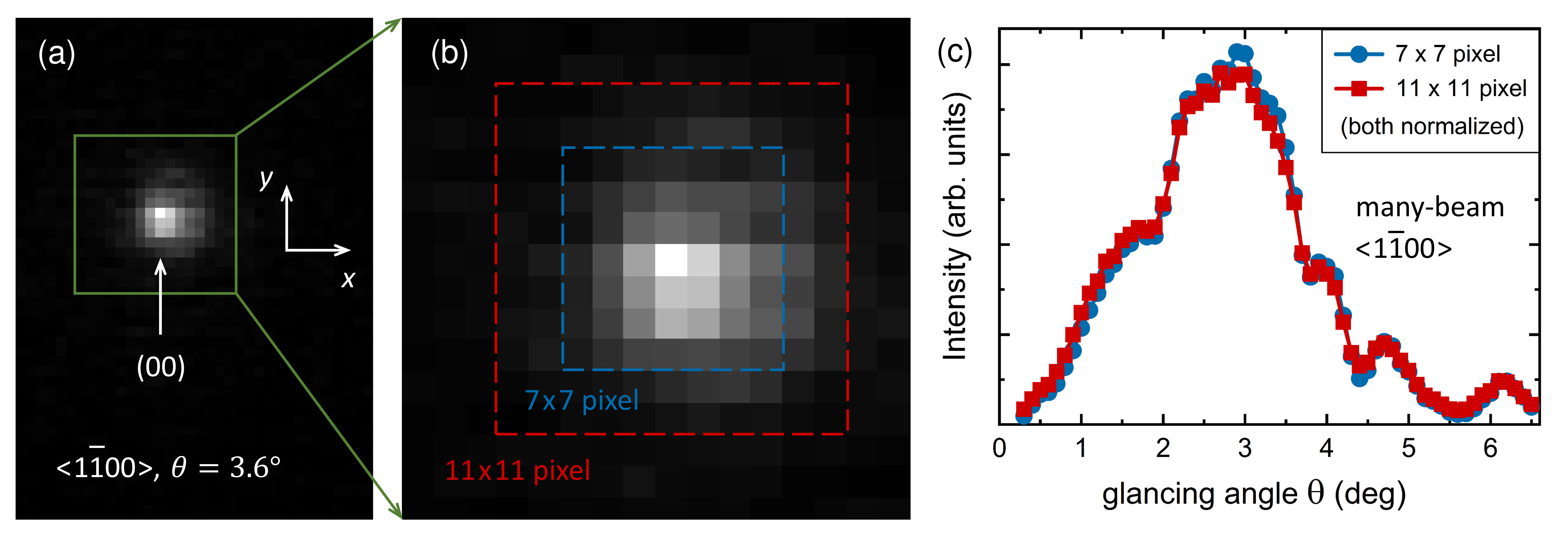}
\setlength\belowcaptionskip{5pt}
\setlength{\abovecaptionskip}{0pt plus 2pt minus 2pt}
\caption{\label{fig:intensity_extraction}Extraction of the (00) spot intensity. (a) Specular spot of QFMLG (25$\thinspace$fps, integration time 40$\thinspace$s). (b) Magnification of the green area in (a), highlighting the intensity of individual camera pixels. The center of the (00) spot was defined by the brightest pixel. The spot intensity was calculated by adding up the brightness values of a $\thinspace 7 \times 7$ or $11 \times 11$ pixel array around the center. \mbox{(c) After} normalization, both approaches yield a similar rocking curve.}
\end{figure}

\mbox{Figure \ref{fig:intensity_extraction}(a)} shows the specular spot of QFMLG recorded along \mbox{$<$1$\bar{1}$00$>$} and for $\theta = 3.6^{\circ}$. For each glancing angle of the rocking scan, $\theta$ was recalculated from the position of the (00) spot to verify the experimental setting and confirm equidistant angular steps.

To obtain the experimental rocking curve, we extracted the intensity of the specular spot from each diffraction pattern for all glancing angles $\theta_i$. This was done in a standardized way to ensure comparability, although the shape and size of the spot varies slightly. For a \mbox{specific $\theta_i$}, the intensity of each pixel was defined as the sum of its brightness values over all frames recorded. Furthermore, we defined the center of the specular spot by the pixel with highest intensity, which is unambiguous since there was no detector saturation. To obtain a measure for the spot intensity, we included several pixels and added up the respective brightness values. Following a well-established procedure, we considered square-shaped pixel arrays of different size around the center, as illustrated in \mbox{figure \ref{fig:intensity_extraction}(b)} for $7 \times 7$ and $11 \times 11$ pixels. 

After normalization, both approaches yield very similar rocking curves, as shown in \mbox{figure \ref{fig:intensity_extraction}(c)}. Including more pixels increases the statistics and can therefore reduce the influence of death pixels or dust on detector and camera. Therefore, we used the $11 \times 11$ data set for the quantitative analysis presented in the main text of this paper. The $7 \times 7$ data set was evaluated to estimate the uncertainty introduced by data processing.

\FloatBarrier

\end{document}